\documentstyle[prl,aps,epsf]{revtex}\def\narrowtext{}\tighten\twocolumn
\input epsf.sty
\begin{document}
\draft

\title{Zn-doping dependence of the wipeout region around Zn in 
YBa$_2$(Cu$_{1-x}$Zn$_x$)$_4$O$_8$}
\author{Y. Itoh,$^{1,2}$ T. Machi,$^1$ N. Watanabe,$^1$ S. Adachi,$^1$ C. Kasai,$^1$ and N. Koshizuka$^1$}

\address{$^1$Superconductivity Research Laboratory,
 International Superconductivity Technology Center,\\
1-10-13 Shinonome, Koto-ku Tokyo 135-0062, Japan \\
$^2$Japan Science and Technology Corporation, Japan}


\address{\begin{minipage}[t]{6.0in} %
\begin{abstract}%
We estimated the spread of the magnetically enhanced regions (wipeout regions) around Zn ions 
in high-$T_c$ superconductors YBa$_2$(Cu$_{1-x}$Zn$_x$)$_4$O$_8$ ($x$=0, 0.005, 0.01, and 0.022) at $T$=4.2 K,
 via the planar $^{63}$Cu nuclear spin-lattice relaxation study with Cu nuclear quadrupole resonance (NQR)
 spin-echo technique. From the analysis of nonexponential planar Cu nuclear spin-lattice 
relaxation curves, we found that the wipeout region per a Zn ion shrinks with Zn doping 
in the superconducting state. The shrinkage is associated with suppression of 
the host antiferromagnetic spin correlation, as a result from a dilution effect 
on the magnetic CuO$_2$ network.   
\typeout{polish abstract} %
\end{abstract}
\end{minipage}} %

\maketitle %
\narrowtext

\section{Introduction}
	Nonexponential relaxation is frequently observed in a wide class of disordered or 
inhomogeneous materials. For nonmagnetic impurity Zn-doped high-$T_c$ superconductors 
YBa$_2$Cu$_3$O$_{7-\delta}$, the nonexponential planar Cu nuclear spin-lattice relaxation curves 
are observed~\cite{Kohori}. A lot of effort has been made to understand how the nonexponential recovery curves 
are described and what is the physics behind it. Recently, it has been shown that 
the magnetic impurity-induced NMR/NQR relaxation theory well reproduces the observed recovery curves 
for impurity-doped high-$T_c$ superconductors ~\cite{Itoh1,Itoh2,Itoh3,Itoh4,Itoh5}, 
which includes the relaxation process 
due to a host Cu spin-fluctuation, that due to an impurity-induced spin-fluctuation, 
and a wipeout effect~\cite{McHenryPRL,McHenryPRB}. The wipeout effect is defined as a loss of NMR/NQR signal 
more than what would be expected from simple dilution effect due to substitution of 
the foreign atoms. The foreign atom is assumed to cause a field gradient sufficiently 
large to shift the resonance frequency of the surrounding nuclei outside the observable range,
 or to cause a local field fluctuation sufficiently large to diminish the NMR/NQR signal
 in the observable time domain. The wipeout effect on Cu NQR spectra has been observed
 in the normal state of YBa$_2$(Cu$_{1-x}$Zn$_x$)$_3$O$_{7-\delta}$~\cite{Yamagata}. 
      
	In this paper, we estimated the wipeout number $N_c$ in the superconducting state 
as a function of Zn content, which could not be estimated from measurements of the Cu NQR spectra,
 via the planar $^{63}$Cu nuclear spin-lattice relaxation study for YBa$_2$(Cu$_{1-x}$Zn$_x$)$_4$O$_8$ (Y124). 
In the light of the impurity-induced NQR relaxation theory with the wipeout effect~\cite{McHenryPRL,McHenryPRB},
 we found shrinkage of the wipeout region around each Zn ion with Zn doping. 

\section{Experimental}
	Powder samples of the Zn-doped Y124 ($x$=0.005, 0.010 and 0.022; $T_c$=68, 56 and 15 K)
 for the Cu NQR experiments are the previously studied ones in Refs. 3-6. Zero-field Cu NQR 
measurements were carried out with a coherent-type pulsed spectrometer. Nuclear spin-lattice 
relaxation was measured by an inversion recovery spin-echo technique, where the $^{63}$Cu(2) 
nuclear spin-echo intensity $M(t)$ was recorded as a function of the time $t$ after an inversion pulse.
 For comparison, the same experiments have been made for the optimally carrier-doped 
YBa$_2$(Cu$_{1-x}$Zn$_x$)$_3$O$_{6.92(5)}$ (Y123) (3$x$=0.02, 0.05 and 0.10; $T_c$=81, 69 and 46 K), which were 
previously synthesized by annealing in reduced oxygen atmosphere in Ref. 10.     
 
\section{Zn-doping dependence of the wipeout number in the superconducting state}
	No appreciable effect of nuclear spin diffusion in space nor in frequency is observed for  Y124 and Y123, 
via (1) the stimulated spin-echo decay study~\cite{Fujiyama}, (2) pulse-strength $H_1$ dependence of the relaxation times, and 
(3) isotope dependence of the recovery curves. From these facts and the practically better fitting results~\cite{Itoh3}, 
one can safely apply the magnetic impurity-induced NQR relaxation theory for Y124 and Y123.
   
Fig. 1 shows the Zn-doping dependence of the experimental recovery curves $p(t)\equiv1-M(t)/M(\infty)$ of 
the planar $^{63}$Cu(2) nuclear magnetization $M(t)$ at 4.2 K. The solid curves are the least-squares fits 
of nonexponential function of,

\vspace{0.3cm}
$p(t)=p(0)exp[-(3t/T_1)_{HOST}-$ \\
\hspace*{1.5cm} $N_c[e^{-3t/t_c}-1+\sqrt{3 \pi t/t_c}erf(\sqrt{3t/t_c})]]$.\hspace{0.5cm} (1)
\vspace{0.1cm}

The notations conform to the previous ones in Refs. 4 and 6. The fit parameters are $p(0)$, 
$(T_1)_{HOST}$, $N_c$ and $t_c$. $(T_1)_{HOST}$ is a Cu nuclear spin-lattice relaxation time due to the host Cu 
electron spin fluctuation, $N_c$ (0 $\leq$ $N_c$ $\leq$ 1) is the wipeout number, $t_c$ is an impurity-induced Cu 
nuclear spin-lattice relaxation time at an exclusion radius $r$=$r_c$ centered around a Zn ion, and $erf$ 
is the error function~\cite{McHenryPRL,McHenryPRB}.

Eq. (1) is based on a {\it minimal model} for dilute alloys, which possesses at least two characteristic 
time constants $(T_1)_{HOST}$ and $t_c$, i.e., the host Cu electron spin correlation via a hyperfine coupling 
and the guest Zn-induced spin correlation via a longitudinal direct dipole coupling (the indirect couplings 
have also possible contributions), respectively. The wipeout effect in Eq. (1) is based on an "{\it all}-{\it or}-{\it nothing}" model. 
The exclusion radius $r_c$ around each Zn is defined such that the nuclei inside the region of radius $r_c$ 
centered around the Zn are unobservable whereas the nuclei outside the region are observable. This model 
seems to be highly simplified, but it is qualitatively consistent with the Zn-induced local moment model 
proposed by $^{89}$Y NMR study above $T_c$~\cite{Mahajan}. 

	Fig. 2(a) shows the Zn-doping dependence of the estimated $N_c$ for Y124 and Y123 at $T$=4.2 K. Here, we 
 assume no Zn(1), i.e., the in-plane Zn concentration $x_{plane}$=2$x$ for Y124 and $x_{plane}$=3$x$/2 for Y123. 
The magnitude of the wipeout number $N_c$ for underdoped Y124 is larger than that for optimally doped Y123. 
The wipeout number $N_c$ increases with Zn doping both for Y124 and Y123. However, the degree of the increase in $N_c$ 
is slower than that expected from disappearance of the Cu nuclei inside a $x_{plane}$-independent neighboring shell by Zn. 
When $I_{jnn}$ is the probability of finding a Zn ion at the $j$-th ($j$=1, 2, 3, 4) nearest-neighboring (nn) shell by Cu(2), 
then $I_{1nn}=I_{2nn}=I_{3nn}$=4$x_{plane}(1-x_{plane})^3$ and $I_{4nn}=8x_{plane}(1-x_{plane})^7$. 
If the Cu(2) nuclei up to the $j$-th nn shells 
by Zn are unobservable, then $N_c^{1nn}=I_{1nn}(x_{plane})$ (dotted curve), 
$N_c^{2nn}=N_c^{1nn}+I_{2nn}(x_{plane})$ 
(short dashed curve), 
$N_c^{3nn}=N_c^{2nn}+I_{3nn}(x_{plane})$ (moderately long dashed curve), and 
$N_c^{4nn}=N_c^{3nn}+I_{4nn}(x_{plane})$ 
(long dashed curve). 
The increase of $N_c$ for Y124 and Y123 is slower than that of $N_c^{4nn}(x_{plane}$) with increasing $x_{plane}$.  
   
 	We approximate $N_c$ by $x_{plane}\pi (r_c/a)^2$, a circle with an effective wipeout radius $r_c$. In Fig. 2(b), 
the estimated wipeout radius $r_c/a$ is shown as a function of the in-plane Zn $x_{plane}$ for Y124 and Y123. 
The wipeout region around each Zn shrinks with Zn doping both for Y124 and Y123. Fig. 3 shows 
the schematic illustration of the shrinkage of the wipeout region in the CuO$_2$ plane. The shaded area 
in the circle indicates the wipeout region around a Zn ion.  

\section{$T$ dependence of the wipeout number} 
	 The estimated values of (1/$TT_1)_{HOST}$, 1/$\tau_1T$ (1/$\tau_1=\pi N_c^2/t_c$ [4, 6-8]), and $N_c$ as functions of $T$ 
for Y124 are shown in Figs. 4(a), (b) and (c), respectively. 

	(a)A small decrease of the normal-state (1/$T_1T)_{HOST}$ with Zn doping is consistent with the previous result 
form the analysis with $N_c$=0 ~\cite{Itoh2,Itoh4,Itoh5}. This can be understood by suppression of the host antiferromagnetic 
correlation as a result from dilution effect of Zn on the magnetic CuO$_2$ network, similarly to the Zn-doped 
parent insulator La$_2$Cu$_{1-x}$Zn$_x$O$_4$~\cite{Carretta}. 

	(b)A small decrease of the Zn-induced relaxation rate 1/$\tau_1T$ just below $T_c$ is observed, which is in contrast 
to the increase of $^7$Li Knight shift in the Li$^+$-doped Y123~\cite{Bobroff}. From a relation of 1/$\tau _1T\propto K\tau$ 
($K$ is the Knight shift at the impurity site, and $\tau$ is the impurity magnetic correlation time), the life time 
$\tau$ of the impurity magnetic correlation decreases just below $T_c$.  

	(c)The value of $N_c$ for $x$=0.01 above $T_c$ is close to unity, so that it seems to be an overestimation. 
This is a shortcoming of the present analysis using Eq. (1), partially because the limitation due to mean impurity 
spacing is not taken into account. However, one can safely conclude that $N_c$ for $x$=0.005 in Fig. 4(c) 
changes smoothly around $T_c$ but does not diverge at $T_c$. This is sharply in contrast to the superconducting 
coherence length $\xi_{SC}$, which must diverge at $T_c$ because the superconducting transition is the second order 
phase transition.   

	Fig. 5 shows the $T$ dependence of the effective wipeout radius $r_c/a$ for the Zn-doped Y124 
($x$=0.005, $x_{plane}$=0.01) 
estimated from a relation of $N_c$=$x_{plane}\pi (r_c/a)^2$. For comparison, various characteristic lengths in units of
an in-plane lattice spacing $a$ are plotted as functions of $T$; the experimental antiferromagnetic correlation 
length $\xi_{AF}$ estimated from the $^{63}$Cu(2) nuclear spin-spin relaxation rate (Gaussian decay rate) 1/$T_{2G}$ ~\cite{Itoh}, 
the superconducting coherence length $\xi_{SC}$=$\xi_{SC}$(0)(1-$T/T_c$)$^{-1/2}$ 
($\xi_{SC}$($T$=0)=4.9$a$ ~\cite{Zech} and $T_c$=68 K) in 
the Ginzburg-Landau (GL) mean-field theory below $T_c$, the superconducting pair correlation length $\xi _{dSC}$ 
in the self-consistent renormalization theory for a two-dimensional $d_{x^2-y^2}$-wave superconducting fluctuation 
model above $T_c$~\cite{Onoda}, and the above $T_c$ superconducting coherence length $\xi_{TDGL}$=$v_F/T$ (the effective Fermi 
velocity $v_F$=1000$a$) in a time-dependent GL theory (the dashed curve)~\cite{Yanase}. The upward triangle 
indicates $\xi_{SC}$=4.9$a$ at $T$=0 extrapolated from an in-plane upper critical field $H_{c2}$ just below $T_c$ [16]. 
The downward triangles indicate an antiferromagnetic correlation length $\xi_{in-gap}$ at $T$=10.5 K of 
Zn-induced in-gap spin fluctuations for the Zn-doped Y123 ($x$=0.02) ~\cite{Sidis}. Obviously, the magnitude 
and the $T$ dependence of $r_c$ for Y124 ($x$=0.005) are similar to $\xi_{in-gap}$ and $\xi_{AF}$ but not to $\xi_{SC}$. 
Thus, one can associate the wipeout region with a locally enhanced antiferrmagnetic correlation region 
around Zn. The wipeout region may correspond to some virtual or real bound state induced by Zn in 
the antiferromagnetic background, so that it reflects the change of the host antiferromagnetic correlation length~\cite{Ohashi}. 
 
\section{Conclusion}
	From the analysis of nonexponential planar Cu nuclear spin-lattice recovery curves, we observed an increase 
of the wipeout number $N_c$ in the CuO$_2$ plane of Zn-doped Y124 and Y123 with Zn doping at 4.2 K. This is similar 
to the increase of the normal-state wipeout number $N_c$ for the Zn-doped Y123~\cite{Yamagata}. However, if $N_c$ is approximated 
by $x_{plane}\pi(r_c/a)^2$, we found that the effective wipeout radius $r_c$ around Zn shrinks 
with Zn doping. We associated the shrinkage and the $T$ dependence of the wipeout region around each Zn with change 
in the host Cu antiferromagnetic correlation. 

\acknowledgements{%
	This work was supported by New Energy and Industrial Technology Development Organization (NEDO) as 
Collaborative Research and Development of Fundamental Technologies for Superconductivity Applications.
} %

\begin{figure}
\caption{
The nonexponential recovery curves $p(t)$ $\equiv1-M(t)/M(\infty)$ of the planar $^{63}$Cu(2) nuclear magnetization $M(t)$
 for Zn-doped Y124 and Y123 at $T$=4.2 K. The solid curves are fitted results based on Eq. (1).
}
\end{figure}

\begin{figure}
\caption{
The wipeout number $N_c$ versus the in-plane Zn content $x_{plane}$ (a), and the effective wipeout radius $r_c$
 versus $x_{plane}$ (b). We assume that Zn predominately substitutes for the planar Cu(2) site, that is,
 $x_{plane}$=2$x$ for Y124 and $x_{plane}$=3$x$/2 for Y123. See the text for the long (4nn), moderately long (3nn),
 short dashed (2nn), and dotted (1nn) curves in (a).  
}
\end{figure} 

\begin{figure}
\caption{
The schematic illustration of the electronic state of the CuO$_2$ plane with Zn.
 The "arrow" represents the enhanced magnetic correlation in the wipeout region around a Zn ion (shaded circles).  
}
\end{figure}

\begin{figure}
\caption{
The estimated (1/$TT_1)_{HOST}$ (a), 1/$\tau_1T$ (1/$\tau_1=\pi N_c^2/t_c$) (b), and $N_c$ (c) as functions of $T$
 for the Zn-doped Y124
 ($x$=0.005 and 0.01). 
}
\end{figure} 

\begin{figure}
\caption{
The $T$ dependence of the effective wipeout radius $r_c$ for the Zn-doped Y124 ($x$=0.005). $\xi_{AF}$, $\xi_{SC}$
 and $\xi_{dSC}$  
in units of an  in-plane lattice spacing $a$ are the antiferromagnetic correlation length [15], 
the superconducting coherence length below $T_c$, and the two-dimensional $d_{x^2-y^2}$-wave
 superconducting pair corrleation length above $T_c$ [17], respectively. The dashed curve is the above $T_c$
 superconducting coherence length $\xi_{TDGL}$ [18]. The upward triangle indicates an extrapolated $\xi_{SC}$=4.9$a$ 
at $T$=0 [16]. The downward triagnles indicate the Zn-induced antiferromagnetic correlation length $\xi_{in-gap}$
 at $T$=10.5 K of Zn-induced in-gap spin fluctuations for the Zn-doped Y123 ($x$=0.02) [19].
}
\end{figure}


\begin{references}
\bibitem{Kohori} Y. Kohori, H. Shibai, Y. Oda, Y. Kitaoka, T. Kohara and K. Asayama, J. Phys. Soc. Jpn. {\bf 57}, 2905 (1988). 
\bibitem{Itoh1}Y. Itoh, T. Machi, N. Watanabe and N. Koshizuka, J. Phys. Soc. Jpn. {\bf 68}, 2914 (1999).   
\bibitem{Itoh2}Y. Itoh, T. Machi and N. Koshizuka, in $Advances in Superconductivity XII$, edited by T. Yamashita and K. Tanabe (Springer-Verlag, Tokyo, 2000), p. 284.
\bibitem{Itoh3}Y. Itoh, T. Machi, N. Watanabe and N. Koshizuka, J. Phys. Soc. Jpn. {\bf        70}, 644 (2001).   
\bibitem{Itoh4}Y. Itoh, T. Machi, N. Watanabe, S. Adachi and N. Koshizuka, J. Phys. Soc. Jpn. {\bf 70}, 1881 (2001).
\bibitem{Itoh5}Y. Itoh, T. Machi, N. Watanabe, and N. Koshizuka, Physica C {\bf 357-360}, 69 (2001).  
\bibitem{McHenryPRL}M. R. McHenry, B. G. Silbernagel and J. H. Wernick, Phys. Rev. Lett. {\bf 27}, 426 (1971).
\bibitem{McHenryPRB}M. R. McHenry, B. G. Silbernagel and J. H. Wernick, Phys. Rev. B {\bf 5}, 2958 (1972).
\bibitem{Yamagata}H. Yamagata, K. Inada and M. Matsumura, Physca C {\bf 185-189}, 1101 (1991).   
\bibitem{Adachi}S. Adachi, C. Kasai, S. Tajima, K. Tanabe, S. Fujihara and T. Kimura, Physica C {\bf 351}, 323 (2001).      
\bibitem{Fujiyama}S. Fujiyama, M. Takigawa, Y. Ueda, T. Suzuki and N. Yamada, Phys. Rev. B  {\bf 60}, 9801 (1999).          
\bibitem{Mahajan}A. V. Mahajan, H. Alloul, G. Collin and J.-F. Marucco, Phys. Rev. Lett. {\bf 72}, 3100 (1994).
\bibitem{Carretta}P. Carretta, A. Rigamoti and R. Sala, Phys. Rev. B  {\bf 55}, 3734 (1997).     
\bibitem{Bobroff}J. Bobroff, H. Alloul, W. A. MacFarlane, P. Mendels, N. Blanchard, G. Collin and J.-F. Marucco, Phys. Rev. Lett. {\bf 86}, 4116 (2001).  
\bibitem{Itoh}Y. Itoh, J. Phs. Soc. Jpn. {\bf 63}, 3522 (1994); {\bf 64}, 684 (1995) [Erratum]. 
\bibitem{Zech}D. Zech, C. Rossel, L. Lesne, H. Keller, S. L. Lee and J. Karpinski, Phys. Rev. B {\bf 54}, 12535 (1996).  
\bibitem{Onoda}S. Onoda and M. Imada, J. Phys. Soc. Jpn. {\bf 68}, 2762 (1999).  
\bibitem{Yanase}Y. Yanase and K. Yamada, J. Phys. Soc. Jpn. {\bf 70}, 1659 (2001).  
\bibitem{Sidis}Y. Sidis, P. Bourges, B. Hennion, L. P. Regnault, R. Villeneuve, G. Collin and J.-F. Marucco, Phys. Rev. B {\bf 53}, 6811 (1996).  
\bibitem{Ohashi}Y. Ohashi, J. Phys. Soc. Jpn. {\bf 70}, 2054 (2001).
\end{references}
\end{document}